\documentclass{article}
\usepackage{spconfa4,amsmath,graphicx}

\usepackage{tikz}
\usetikzlibrary{calc}
\usetikzlibrary{shapes.geometric}
\usetikzlibrary{math} 
\usepackage{multirow}
\usepackage{amsmath}
\usepackage{cite}
\usepackage{amsfonts}
\usepackage[subtle]{savetrees}
\usepackage{caption}
\usepackage{xcolor}
\usepackage{hhline}
\usepackage[normalem]{ulem}
\usepackage{soul}
\usepackage{xcolor}

\title{Room Impulse Response prototyping using receiver distance estimations for high quality room equalisation algorithms}
\threeauthors
 {James Brooks-Park $^1$ \sthanks{This project has received funding from the European Union's Horizon 2020 research and innovation programme under the Marie Skłodowska-Curie grant agreement No 956369.}, Steven van de Par $^1$}
	{$^1$Carl von Ossietzky Universität Oldenburg\\
	Acoustics Group\\
	Oldenburg, Germany}
 {Jan Østergaard $^2$}
	{$^2$Aalborg University\\
	Dept. of Electronic Systems,\\
	Aalborg, Denmark}
  {Søren Bech $^{2,3}$, Martin Bo Møller $^3$}
	{$^3$Bang \& Olufsen a/s\\
	Struer, Denmark\\
  }

\def\centerarc[#1](#2)(#3:#4:#5)% Syntax: [draw options] (center) (initial angle:final angle:radius)
    { \draw[#1] ($(#2)+({#5*cos(#3)},{#5*sin(#3)})$) arc (#3:#4:#5); }

\def\Waves[#1](#2)(#3:#4);{
  \begin{scope}
    \centerarc[#1](#2)(#3:#4:0.2)
    \centerarc[#1](#2)(#3:#4:0.4)
    \centerarc[#1](#2)(#3:#4:0.6)
  \end{scope}
}
\def\speaker(#1)(#2)(#3);{
    \begin{scope}[shift={(#1)},rotate={#2},scale={#3}]
        \draw[line width = .1mm] (-0.8,-0.40) rectangle (-0.35,0.40);
        \draw[line width = .1mm] (-0.35,0.40) -- (0,0.65) -- (0,-0.65) -- (-0.35,-0.40);
        \Waves[gray,thick](0,0)(60:0);
        \Waves[gray,thick](0,0)(360:300);
    \end{scope}
}

\begin{document}
%\ninept
%
\maketitle

\begin{abstract}
\vspace*{-0.2cm}
Room equalisation aims to increase the quality of loudspeaker reproduction in reverberant environments, compensating for colouration caused by imperfect room reflections and frequency dependant loudspeaker directivity. A common technique in the field of room equalisation, is to invert a prototype Room Impulse Response (RIR). Rather than inverting a single RIR at the listening position, a prototype response is composed of several responses distributed around the listening area. This paper proposes a method of impulse response prototyping, using estimated receiver positions, to form a weighted average prototype response. A method of receiver distance estimation is described, supporting the implementation of the prototype RIR. The proposed prototyping method is compared to other methods by measuring their post equalisation spectral deviation at several positions in a simulated room.
\end{abstract}

% __________________________________________________________________________Introduction___________________________________________________________________________________________
\vspace*{-.4cm}
\section{Introduction}
\vspace*{-.3cm}
\label{sec:intro}

Room equalisation has been a key area of research and development over the last 45 years, with many applications ranging from cinema audio systems, home HI-FI products, and more recently, network audio devices. There are many methods of tackling room equalisation, each with their associated drawbacks and benefits \cite{cecchi_room_2017}. One of the challenges that all applications face, is the variability of RIR's within a room / listening area \cite{mourjopoulos_variation_1985}. Many factors affect the composition of a RIR, the source / receiver position and orientation, room temperature, and furniture positioning. In many situations these components are not static, decreasing the accuracy of RIR measurements with even the smallest of changes to the room's composition.
% \begin{figure}\label{RIR}
%     \begin{center}
%         \includegraphics[width=\linewidth]{./Figures/FrequencyPlot.pdf}
%     \end{center}
%     \vspace{-.5cm}
%     \caption{Comparison of individual spatially distributed RIR's (in Grey) to the proposed weighted average prototype transfer function (in black)}
%     \vspace{-.6cm}
% \end{figure}

Room equalisation can offer a dramatic improvement to the listening experience of an audio reproduction system. However, due to the static nature of an equalisation filter, particular considerations should be made to design a robust filter to accommodate for small changes to the RIR. Instead of using a single RIR to design a filter, several RIRs can be combined, creating a prototype RIR containing information about a wider area, increasing its positional (spatial) robustness and reducing the chance of over-fitting\cite{norcross_inverse_2006}.

There are many prototype impulse response designs that have been proposed, the variability of zeros of the transfer functions in the complex plane across several receiver positions can be analysed, determining which can or cannot be inverted \cite{brannnmark_robust_2008}. Clusters of RIRs can be formed based on their spectral Euclidean distance from one another, these clusters can be averaged and used as a prototype RIR\cite{bharitkar_cluster_2001}. Perhaps the most common method of generating a prototype impulse response, is to take the mean average power spectrum across several positions in a listening area, reducing the influence of spectral features that are not consistent across all measurement positions. It has been shown in \cite{pedersen_sampling_2007} that increasing the number of measurements in a room, improves the accuracy of the averaged power response at low frequencies (below 1000 Hz). This basic methodology is often used and has been expanded upon to be the core of many prototyping methods \cite{elliott_multiple-point_nodate,bharitkar_robustness_2002,abildgaard2006loudspeaker,jin_acoustic_2022,bharitkar_robustness_2004}.

Whilst an average prototype response may represent the sound field across a listening area, there are further and often neglected considerations that could be made. Traditional loudspeakers typically feature a flat on-axis response, and a directivity pattern characterized by highly directive high frequencies and nearly omnidirectional low frequencies. To get the most accurate reproduction experience, the listener would aim to be positioned on-axis horizontally and vertically to the loudspeaker. In the case of stereophonic reproduction, the listener would also aim to accurately reproduce the stereo image, this can only be achieved with the correct loudspeaker listener configuration. With larger loudspeaker configurations, the spatial image becomes increasingly more complex, escalating the demand for accurate spatial reproduction. The combination of a loudspeaker's directivity pattern and the reproduced media's spatial image, demands an optimal listening position \cite{british-standard-institution_multichannel_2022}. 

Established prototyping methods do not consider the directivity characteristics of a loudspeaker, where spectral information from outside of the optimal listening area causes the prototype response to have perceivably different spectral characteristics to the optimal listening position. Given the directive nature of high frequencies, a high amount of variation is to be expected between receiver positions. This could produce a prototype response with less high frequency energy compared to the optimal position, resulting in a room compensation filter that overcompensates at higher frequencies. Whilst this may be initially perceived as brilliance, or a sense of clarity, this is in fact an inaccuracy.

This paper proposes a prototype response formed of several RIRs weighted based upon their distance from the optimal listening position and further weighted by frequency. The motivation behind this work is to design a prototype response that considers the variability of RIRs, as well as acknowledging the importance of the optimal listening position, defined by the loudspeaker listener configuration and directivity characteristics.

This paper first introduces a method of receiver distance estimation, knowing only the distance between two sources and the RIRs at several positions in the listening area. These estimated distances are used to form the weights of a prototype response for room equalisation algorithms. Results from both methods are evaluated via a simulation study.

% __________________________________________________________________________Methodology___________________________________________________________________________________________
\vspace*{-.3cm}
\section{Methodology}
\vspace*{-.2cm}
\label{sec:method}
% \vspace*{-.1cm}
\subsection{Distance Calculation}\label{Distance}
% \vspace*{-.1cm}

Knowledge of the distance between receiver positions can be useful when trying to understand the characteristics of a sound field. However, measuring these distances whilst recording RIRs, can be a time consuming and potentially inaccurate process. This section proposes a method where distances between receiver positions are estimated using RIRs, from two sources with only the distance between the sources requiring a physical measurement.

\begin{figure}[ht]
    \begin{center}
        \begin{tikzpicture}
            % Circles/Text
            \draw[font = \tiny,fill = gray] (-2.5,0) circle (0.2); 
            \node[font = \large] at (-2.5,0.4) {S1};
            \draw[font = \tiny,fill = gray] (2.5,0) circle (0.2); 
            \node[font = \large] at (2.5,0.4) {S2};
            \draw[font = \tiny,fill = gray] (1.5,-2) circle (0.2); 
            \node[font = \large] at (1.3,-2.4) {R};

            % Joining Lines
            \draw[line width=.5mm] (-2.2,0) -- (2.2,0);  
            \node[font = \large] at (0,0.4) {b};
            \draw[line width=.5mm,color = lightgray] (-2.5,-.3) -- (1.2,-2);
            \node[font = \large,color = lightgray] at (-1,-1.3) {P1/c};
            \draw[line width=.5mm,color = lightgray] (2.5,-.3) -- (1.7,-1.8);
            \node[font = \large,color = lightgray] at (2.45,-1.3) {P2/a};  
                
            % Coordinate Lines
            \draw[<->,line width = 2] (1.5,-2.4)--(2.5,-2.4);
            \node[font = \large] at (2,-2.6) {x};
            \draw[<->,line width = 2] (2.9,-2)--(2.9,0);
            \node[font = \large] at (3.1,-1) {y};
            
            % Sub triangle lines
            \draw[line width=.5mm,color = darkgray] (1.5,-1.65) -- (1.5,-.2) -- (2.3,-.2) -- cycle;

        \end{tikzpicture}
        
    \end{center}
    \vspace*{-.7cm}
    \caption{Example of a two source and receiver triangle (Black/Light Grey), and sub triangle (Grey) for acoustic receiver coordinate estimation}
    \vspace*{-.5cm}
    \label{fig:TriangleDiagram}
\end{figure}
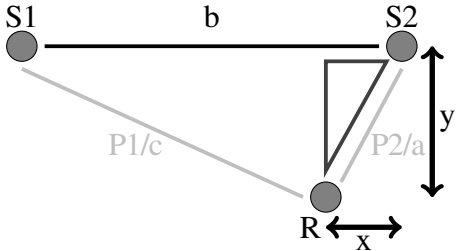

The proposed method treats the two sources and the receiver as a triangle (light grey), and later a sub triangle (grey) as shown in Figure \ref{fig:TriangleDiagram}, the geometric features of the sub triangle are used to create coordinates for each receiver position. This proposed method assumes that all sources have either +/- y coordinates, meaning all positions must be in front of (or behind) the loudspeakers.

To estimate the geometrical features of the triangles, some initial information must be obtained. The distance between the two sources (b) is known, and the distance between source and receiver can be estimated with the Time Of Arrival (TOA). The TOA can be estimated in various ways, the simplest finding the first sample that is greater than $50\%$ of the maximum value of the RIR. This method acknowledges that the direct sound may not be the largest peak in the RIR, which is possible when a receiver is near a reflective boundary. Knowing the sample where the direct sound arrives, the distance can be calculated using the sampling frequency and the speed of sound.

The height of the triangle is the y coordinate of the receiver position, one can simply find this by first calculating the semi-perimeter of the triangle, with this Herons formula can be used to find the area of the triangle. With the area of the triangle and the measured base between the two sources, the height (y coordinate) can be calculated.

With the height of the receiver and source triangle, the base of the sub triangle (Shown in grey in Figure \ref*{fig:TriangleDiagram}) is the x coordinate of the receiver. It is important to note that if the receiver position is outside of the sources' x coordinate, the coordinate zero source must be selected so that both the receiver positions are on the same side of the receiver. For example, in Figure \ref*{fig:TriangleDiagram}, $P1$ is $c$ as it's distance is greater than $P2$.

With the estimated receiver coordinates the distance between two receivers ($z$) can be calculated

\begin{equation}
    z = \sqrt{(y_{opt}-y_{ext})^2+(x_{opt}-x_{ext})^2}\;\;.
\end{equation}

Where $y_{opt}$ and  $x_{opt}$ are the estimated coordinates for the optimal listening position, and $y_{ext}$ and $x_{ext}$ are the estimated coordinates for the external receiver position.

% \vspace*{-.1cm}
\subsection{Global Transfer Calculation}\label{sec:GTF}
% \vspace*{-.1cm}

\begin{figure}
    \centering
    \includegraphics[width = \linewidth]{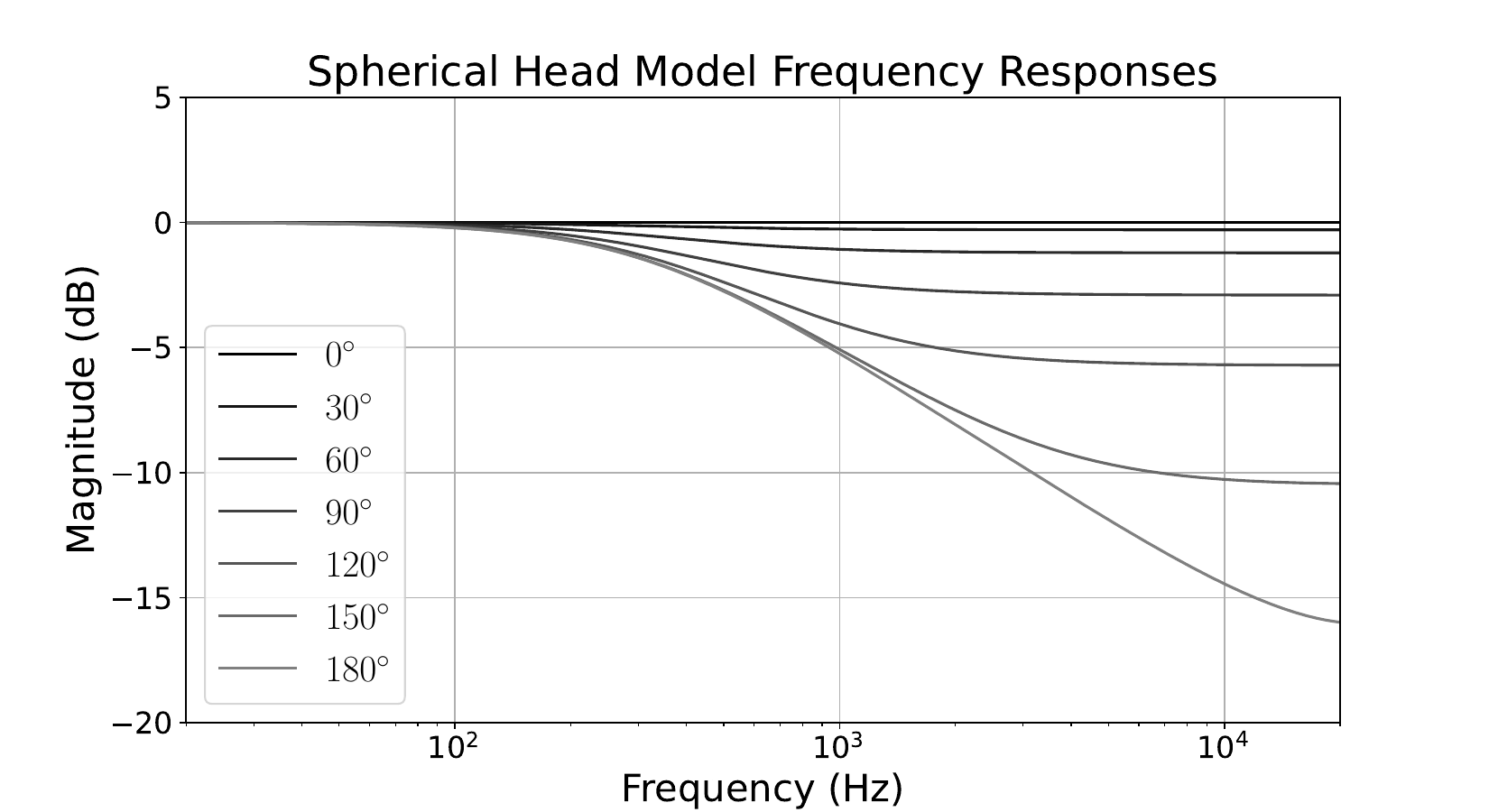}
    \vspace*{-.7cm}
    \caption{Spherical Head Model frequency responses with varying azimuth angles, normalised so the on-axis response is flat.}
    \label{SHM_Fig}
    \vspace*{-.6cm}
\end{figure}

In the proposed prototype response, estimated receiver distances are used to proportionally weight RIRs so that responses positioned further from the optimal listening position are assigned smaller weights. A secondary weighting is applied to each receiver position, based on the angle ($\theta$) between the on-axis loudspeaker and receiver. Given the frequency dependant directivity characteristics of a loudspeaker (modelled in Figure \ref*{SHM_Fig}), lower frequencies are more generalizable across a larger area, compared to higher frequencies that have a high variance dependant on the azimuth and elevation angle. The generalizable frequency response of a loudspeaker is modelled as a spherical head model (SHM), which has been proposed to model the directivity characteristics of a loudspeaker in room acoustic simulations \cite{ewert_computationally_2021}.

The weighting for each RIR $W(\omega, z, \theta)$ is a product of the frequency weighting $W_\text{freq}(\omega, \theta)$ and distance weighting $W_\text{dist}(z)$

\begin{equation}
	W(\omega, z, \theta) = W_\text{dist}(z) W_\text{freq}(\omega, \theta)\;.
\end{equation}

The distance $z$ is the estimated distance between the optimal listening position and the receiver in meters, if the receiver is at optimal listening position, $z = 0 m$. To create the distance weighting $W_\text{dist}(z)$, $z$ is offset by $\delta$ and exponentially modified by $\zeta$, the combination of these modifications assigns lower weights to positions that are further from the optimal listening position

\begin{equation}
    W_\text{dist} = (z+\delta)^\zeta\;,
    \quad
    \zeta \in [-1,0] \;.
\end{equation}

Decreasing $\delta$ increases the initial difference between the optimal receiver position's weight, and the distributed receiver weights. Whilst $\zeta$ controls the rate of decay over distance, where a larger value increases the rate of decay. For the results presented in Section \ref*{sec:results}, $\delta$ is set to 0.01 and $\zeta$ is -0.4 as these parameters have been determined to be optimal weights through a parameter sweep. 

The frequency weighting $W_\text{freq}(\omega, \theta)$ assigns a weight based on the receivers azimuth angle to the loudspeaker, using the generalised loudspeaker directivity response at the vertex ($\theta$) between sides $a$ and $b$ 
\vspace*{-.3cm}

\begin{equation}
    \theta = \arccos((a^2+b^2+c^2)/2ab) \;.
\end{equation}
\vspace*{-.4cm}

The method of modelling the SHM frequency directivity characteristics proposed in \cite{brown_structural_1998}, and the implementation from \cite{zolzer2002dafx} is used to create $W_\text{freq}(\omega, \theta)$. The responses are normalised so that the on-axis response is flat, as shown in Figure \ref*{SHM_Fig}.

The weighting function $W(\omega, z, \theta)$ is normalized to have a maximum value at unity. The proposed prototype response is formed of a weighted average of several RIR power responses. A frequency dependant weighting factor $W(\omega, z_r, \theta_r)$ is applied to each receiver position, the average of these responses forms the prototype RIR, as shown in Equation (\ref*{averageeq})

\vspace*{-.2cm}
\begin{equation}
	H_\text{p}(\omega) = \frac{1}{R} \sum_{r=0}^{R-1} \sqrt{ |H_r(\omega) W(\omega, z_r, \theta_r) }\;,
    \label{averageeq}
\end{equation}
\vspace*{-.2cm}

where $z_r$ and $\theta_r$ denote the distance and angle of the $r$th receiver position, respectively.

% __________________________________________________________________________Results___________________________________________________________________________________________
\vspace*{-.4cm}
\section{Results}
\vspace*{-.4cm}
\label{sec:results}
\vspace*{-.2cm}
\subsection{Simulations}
\vspace*{-.2cm}

The proposed prototyping method is compared to an unweighted average and a single RIR, using RIRs generated by the room acoustic simulator RAZR \cite{wendt2014computationally}. To get enough variability both across the frequency range and receiver positions, the SHM is used as a frequency dependant source directivity model, as this has been shown to be a reasonable approximation of a general loudspeaker directivity pattern \cite{ewert_computationally_2021}.

Simulations are run for four different rectangular room geometries, the positions of the sources and receivers, shown in Figure \ref*{RoomDiagram}, are scaled proportionally to the varying room dimensions. RIRs are generated for the optional listening position and 6 external receiver positions are randomly generated within the listening area (displayed as the grey box in Figure \ref*{RoomDiagram}), these randomly positioned receivers are used to form the prototype response. For each room geometry, the simulation is repeated 5 times to vary the external receiver positions. To analyse the effect of the prototype responses, 100 RIRs are generated forming a uniform grid across the listening area. 

\begin{figure}
    \caption*{Room Diagram}
    \vspace*{-0.7cm}
    \begin{center}
        \scalebox{.6}{
            \begin{tikzpicture}
                % Room
                \draw[line width = .6] (0,0) -- (8,0) -- (8,10) -- (0,10) -- cycle;
                % Source Positions
                \draw[fill = black] (2,7) circle (.1);
                \draw[fill = black] (6,7) circle (.1);

                \node[] at (2,7.5) {S1};
                \node[] at (6,7.5) {S2};

                % Optimal receiver position
                \draw[fill = black] (4,3.536) circle (.1);
                \node[] at (4,3) {$\text{R}_\text{opt}$};

                % Equilateral triangle
                \draw[dashed] (2,7) -- (6,7) -- (4,3.536) -- cycle;
                \node[] at (2.65,5.3) {c};
                \node[] at (5.35,5.3) {a};
                \node[] at (4,7.2) {b};

                % Listening area
                \draw[fill = lightgray,opacity = .2] (2,4.5) -- (6,4.5) -- (6,2.5) -- (2,2.5) -- cycle;

                % 2m Lines
                \draw[dotted] (2,4.5) -- (-1,4.5);
                \draw[dotted] (2,2.5) -- (-1,2.5);

                \draw[<->] (-1,4.4) -- (-1,2.6);
                \node[] at (-.5,3.536) {2m};

                % 30% line
                \draw[dotted] (-1,7) -- (9,7);
                \node[] at (-.5,7.2) {70\%};
                
                % 25% and 75% lines
                \draw[dotted] (2,10.7) -- (2,-.5);
                \draw[dotted] (6,10.7) -- (6,-.5);

                \node[] at (2.5,10.4) {25\%};
                \node[] at (6.5,10.4) {75\%};

                %  W & L
                \node[font = \Large] at (4,-.5) {W};
                \node[font = \Large] at (8.5,5) {L};

            \end{tikzpicture}
        }
    \end{center}
    \vspace*{-.5cm}
    \caption{Simulation room design, sources are positioned at 25\% and 75\% of the room's width on the $x$ axis and 70\% of the length on the $y$ axis, the optimal receiver is positioned so that $a = b = c$, with the loudspeakers angled to face the optimal listening position. The height of the sources is 1.2 across all simulations. The listening area (grey) is 2m in length (where boundary limits allow) and the width is determined by the sources $x$ coordinates. The variable room dimensions (W, L, H) are: (5, 7, 3), (6, 8, 3.5), (7, 9, 3.5), (8,6,3.5).}
    \label{RoomDiagram}
    \vspace*{-.9cm}
\end{figure}
\vspace*{-.4cm}
\subsection{Receiver Distance}
\vspace*{-.2cm}

Comparing the receiver distances estimated using the method proposed in Section \ref*{Distance}, to geometrically calculated distances, the proposed method has a mean average error of 4.6mm, across 336 randomly distributed receiver positions within areas ranging from $5 \text{ to }7 m^2$. With a minimum and maximum error of 0.004mm and 14.9mm respectively, and a standard deviation of 3.45mm.

\vspace*{-.4cm}
\subsection{Prototype Impulse Response}
\vspace*{-.2cm}

The original motivation for a room equalisation prototype impulse response, is to both increase the filters robustness to the listeners position, and to accommodate for small changes to the rooms RIR. The method proposed in this paper introduces a secondary aim, ensuring a low level of spectral deviation at the optimal position, whilst maintaining positional robustness across the listening area. The spectral deviation metric is used to objectively measure the performance of the room equalisation filters, when a spectrally flat target function is selected, a lower spectral deviation value would indicate an increase in performance. For the purpose of analysing the performance of the proposed approach, RIRs and prototype responses are inverted using the filter $H(\omega)^{-1}$, described in \cite{norcross_inverse_2006}. 

\vspace*{-0.2cm}
\begin{equation}\label{Inverse}
    H(\omega)^{-1} =  \frac{H^*(\omega)}{\lvert H(\omega) \rvert^2 + \beta \lvert H(\omega)\rvert^2}
\end{equation}

Where $H$ is the RIR or prototype RIR to be inverted, with the regularisation factor $\beta$, set to 0.01 for this simulation campaign.

The calculated filters are applied to loudspeakers in the previously described simulations, and the spectral deviation  is calculated for each RIR position before and after equalisation 

\vspace*{-0.52cm}
\begin{equation}\label{specdev1}
    S_D = \sqrt{\frac{1}{Q_h-Q_l-1}\sum_{i=Q_l}^{Q_h}(20\log_{10}\lvert Y(\omega_i)\rvert - D)^2} \; , 
\end{equation}
\vspace*{-0.6cm}
where
\begin{equation}\label{specdev2}
    D = \frac{1}{Q_h-Q_l-1}\sum_{i=Q_l}^{Q_h}20\log_{10}\lvert Y(\omega_i)\rvert \;\;.
\end{equation}
%\vspace*{-0.3cm}

The constants $Q_h$ and $Q_l$ are the upper and lower frequency limits for the spectral deviation calculation and $Y(\omega)$ is the frequency domain magnitude response.

Table \ref*{PrototypeResults} presents the simulated spectral deviation results for the proposed prototype response (Weighted) and for comparison, no equalisation (No EQ), a single measurement equalisation (Local), and an unweighted average (Unweighted). The low frequency band ranges from 50Hz to 2kHz and the high frequency band extends from 2kHz to 20kHz. These bands correlate to splitting the basilar membrane in half with regards to its frequency distribution, relating directly to how humans perceive sound across the audible frequency range.
% \vspace*{-.2cm}
\begin{table}[t]
    \centering
    % \vspace*{-.2cm}

    \caption*{Spectral deviation results (dB)}
    \vspace*{-.2cm}
    \tabcolsep=0.05cm
    \begin{tabular}{c|c|c|c|c|c}
    \small Pos. &\small Band   & \small No EQ                         &\small Local                           &\small Unweighted                 &\small Weighted         \\\hline
    \small Opt. &\small Total  &\small2.04{\footnotesize[1.99,2.1]}   &\small0.45{\footnotesize[0.45,0.46]}    &\small1.7{\footnotesize [1.66,1.73]}   &\small1.5{\footnotesize[1.48,1.54]}\\
    \small      &\small Low    &\small3.65{\footnotesize[3.59,3.71]}  &\small0.37{\footnotesize[0.36,0.39]}    &\small2.69{\footnotesize[2.61,2.77]}   &\small2.43{\footnotesize[2.37,2.49]}\\
                &\small High   &\small1.76{\footnotesize[1.7,1.82]}   &\small0.43{\footnotesize[0.42,0.44]}    &\small1.5{\footnotesize[1.47,1.54]}    &\small1.33{\footnotesize[1.3,1.37]}\\\hline \hline
    \small Avg. &\small Total  &\small2.08{\footnotesize[2.07,2.08]}  &\small3.07{\footnotesize[3.07,3.08]}    &\small1.85{\footnotesize[1.84,1.86]}   &\small1.87{\footnotesize[1.87,1.88]}\\
    \small      &\small Low    &\small3.52{\footnotesize[3.51,3.53]}  &\small4.46{\footnotesize[4.45,4.48]}    &\small2.78{\footnotesize[2.77,2.78]}   &\small2.81{\footnotesize[2.81,2.84]}\\
                &\small High   &\small1.8{\footnotesize[1.8,1.81]}    &\small2.73{\footnotesize[2.72,2.74]}    &\small1.67{\footnotesize[1.66,1.67]}   &\small1.67{\footnotesize[1.67,1.67]}
    \end{tabular}
    \vspace*{0.2cm}
    \caption{Post equalisation spectral deviation results, simulating different receiver positions and prototyping methods. Mean values are presented, with 95\% confidence intervals in square brackets. Pos. = Position, either at the optimal listening position (opt.) or averaged across the listening area (Avg.)}
    \label{PrototypeResults}
    \vspace*{-.9cm}

\end{table}

The number of external receiver positions used to create the proposed prototype response affects the spectral deviation performance at the optimal position and across the listening area. When analysing the performance between the number of external receivers (4, 6, 8 \& 10), fewer receiver positions lead to lower levels of spectral deviation at the optimal listening position, as the optimal receiver position will be weighted higher. As one may expect, the average performance across the listening area is increased with the number of external receiver positions, as more information is contained in the prototype response across the evaluation area. For the results displayed in Table \ref*{PrototypeResults}, 6 external receiver positions are used, as this provided the lowest total spectral deviation for the reference and average results.
\begin{table}[tp]
    \centering
    \begin{tabular}{c|c|c|c|c|c}
        $\delta$ & 1 & 0.1 & 0.01 & 0.001& 0.0001\\\hline
        Ref. & 1.96 & 1.87 & 1.76 & 1.76 & 2.12\\
        Avg. & 2.07 & 2.08 & 2.12 & 2.28 & 2.75\\
    \end{tabular}
    \vspace*{-0.2cm}

    \caption{Spectral deviation results for varying $\delta$ parameter values, with a $\zeta$ value of -0.4}
    \label{ExponentResults}
    \vspace*{-0.4cm}

\end{table}
% \vspace*{-0.4cm}
\begin{table}[tp]
    \centering
    \begin{tabular}{c|c|c|c|c|c}
        $\zeta$ & -0.1 & -0.2 & -0.4 & -0.6& -0.8\\\hline
        Ref. & 1.93 & 1.87 & 1.75 & 1.74 & 2.11\\
        Avg. & 2.07 & 2.08 & 2.12 & 2.27 & 2.72\\
    \end{tabular}
    \vspace*{-0.3cm}

    \caption{Spectral deviation results for varying $\zeta$ parameter values,with a $\delta$ value of 0.01}
    \label{OffsetResults}
    \vspace*{-0.7cm}
\end{table}

The exponent ($\zeta$) and offset ($\delta$) parameters used for the presented simulations were chosen as they resulted in the lowest total spectral deviation for the reference and average results. The results from the parameter sweep (Shown in Tables \ref*{ExponentResults} and \ref*{OffsetResults}) displayed a linear relationship between the average spectral deviation across the listening area, where the proposed prototype response increased in performance as the offset parameter was increased, and the exponent parameter was decreased. Whilst one may expect the results at the optimal listening position to display a linear relationship in the opposite direction, this was not the case. Both parameters exhibit the shape of a quadratic function for the optimal position results.

To analyse the performance of the proposed method, a comparison is made to an unweighted average prototype. Comparing the spectral deviation results at both the optimal position and across the listening area, an assessment can be made as to whether the proposed method increases the performance at the optimal position, whilst not degrading the performance at other positions within the listening area. At the optimal position, there is a 0.2 dB decrease in spectral deviation, with an average increase of just 0.02 dB across the listening area, demonstrating the proposed method achieving its original aim. This is statistically supported by a pooled variance estimate T-test, where the comparison distributions are tested, resulting in a p value less than 0.05, showing the differences between the discussed results to be statistically significant.

% __________________________________________________________________________Conclusion___________________________________________________________________________________________
\vspace*{-.5cm}
\section{Conclusion}
\vspace*{-.3cm}
\label{sec:conclusion}

When designing a room equalisation filter for a loudspeaker configuration that demands an optimal listening position, the proposed prototype response promises to improve the accuracy of the equalisation filter at the optimal position whilst maintaining spatial robustness. The results presented in Section \ref*{sec:results} show the proposed approach outperforming an unweighted average prototype at the optimal position, whilst not degrading the performance across the listening area. Supported by an acoustic receiver distance estimation method that is more time-efficient and less prone to human error in comparison to manual measurements. The results presented in Tables \ref*{ExponentResults} and \ref*{OffsetResults} depict tuneable parameters that can be optimised by an end user to increase the spatial robustness of their room equalisation filter. Whilst the objective results presented in this paper may indicate the performance of the proposed approach, a subjective listening experiment would need to be performed to analyse the perceptual benefits of the discussed results.
\bibliographystyle{IEEEbib}
\bibliography{refs}

\end{document}